\documentclass[runningheads,a4paper]{llncs}
\usepackage{graphicx}
\usepackage{amsmath}
\usepackage{amssymb} 
\usepackage{xspace}

\usepackage{array}

\usepackage[linesnumbered,ruled]{algorithm2e}

\usepackage{color}
\usepackage{xcolor}

\newif\ifArx
\Arxfalse 
\newcommand{\props}{P}
\newcommand{\model}{\mathcal{M}}

\newcommand{\closure}{\mathcal{C}}

\newcommand{\dist}[1]{\mathcal{D}^{#1}} 

\newcommand{\reals}{\mathbb{R}}
\newcommand{\nats}{\mathbb{N}}

\newcommand{\arel}{\mathcal{R}}

\newcommand{\SLCS}{SLCS\xspace}
\newcommand{\SLCSMI}{ImgQL\xspace}

\newcommand{\imgqool}{{\ttfamily{VoxLogicA}}\xspace}

\newcommand{\lnear}{\mathcal{N}}
\newcommand{\lsurr}{\mathcal{S}}

\ifArx
\SetAlFnt{\small}
\SetAlCapFnt{\small}
\SetAlCapNameFnt{\small}
\else
\SetAlFnt{\small}
\SetAlCapFnt{\small}
\SetAlCapNameFnt{\small}  
\fi
\SetAlCapHSkip{0pt}
\IncMargin{-\parindent}

\newcommand{\imgImport}{{\ttfamily{import}}\xspace}

\newcommand{\msep}{\,\mid\,}
\newcommand{\form}{{\bf F}}

\colorlet{VINC}{orange}
\colorlet{GINA}{blue}

\newcolumntype{L}{>{$}l<{$}}

\usepackage{subfig}
\usepackage{tikz}
\usepackage{latexsym}
\definecolor{darkgreen}{RGB}{30,120,30}

\newcommand{\RED}[1]{\textcolor{red}{#1}}
\newcommand{\BROWN}[1]{\textcolor{brown}{#1}}

\newcommand{\GREEN}[1]{\textcolor{darkgreen}{#1}}
\newcounter{dgnot}
\newenvironment{dgnot}[1][]{\refstepcounter{dgnot}\par\medskip
   \noindent \textbf{\RED{NfD~\thedgnot.}  #1} \rmfamily}{\medskip}

\newcommand{\ed}{\hfill$\bullet$}

\newcommand{\SET}[1]{\{#1\}}
\newcommand{\ZET}[2]{\SET{#1 \mid #2}}

\newcommand{\aeval}{{\cal A}}
\newcommand{\peval}{{\cal V}}

\newcommand{\lneg}{\neg}
\newcommand{\lfalse}{\bot}
\newcommand{\ltrue}{\top}

\newcommand{\sem}[1]{[\![#1]\!]}

\newcommand{\lssim}[7]{\triangle\!\!\!{\scriptstyle \triangle}_{#1}
{\tiny 
\left[
\begin{array}{ccc}
#2 & #3 & #4\\
#5 & #6 & #7
\end{array}
\right]
}}

\newcommand{\mkhis}{{\cal H}}
\newcommand{\mean}[1]{\overline{#1}}
\newcommand{\cc}{\mathbf{r}}

\renewcommand{\form}{\Phi}
\renewcommand{\arel}{R}

\usepackage{doi}
\usepackage{url}

\newcommand{\MAGENTA}[1]{\textcolor{magenta}{#1}}

\newcounter{mknot}
\newenvironment{mknot}[1][]{\refstepcounter{mknot}\par\medskip
   \noindent \textbf{\MAGENTA{NfM~\themknot.}  #1} \rmfamily}{\medskip}

\newcounter{vcnot}
\newenvironment{vcnot}[1][]{\refstepcounter{vcnot}\par\medskip
   \noindent \textbf{\BROWN{NfV~\thevcnot.}  #1} \rmfamily}{\medskip}

\newcounter{gbnot}
\newenvironment{gbnot}[1][]{\refstepcounter{gbnot}\par\medskip
   \noindent \textbf{\GREEN{NfG~\thegbnot.}  #1} \rmfamily}{\medskip}

\providecommand{\url}[1]{{#1}}

\newcommand{\old}[1]{}

\begin{document}

\mainmatter


\ifArx
\title{Using Spatial Logic and Model Checking for\\ Nevus Segmentation} 
\else
\title{Using Spatial Logic and Model Checking for\\ Nevus Segmentation\thanks{Part of this work has been developed in the context of the Italian MIUR-PRIN 2017 project IT MaTTerS: Methods and Tools for Trustworthy Smart Systems". The names of the authors are listed in alphabetical order. The major contributions to the development of the specification in ImgQL are those of G. Broccia and M. Massink, who also have been taking care running it on the datasets and collecting the results. D. Latella and V. Ciancia also contributed to the development of the specification. V. Ciancia developed \imgqool and G. Belmonte proposed the particular texture analysis approach. All authors contributed to drafting and finalising the paper. }}
\fi
\titlerunning{SL\&MC for Nevus Segmentation}

\author{Gina Belmonte\inst{1} \and Giovanna Broccia\inst{2} \and Vincenzo Ciancia\inst{2} \and\\ Diego Latella\inst{2} \and Mieke~Massink\inst{2}}
\institute{Azienda Toscana Nord Ovest S. C. Fisica Sanitaria Nord, Lucca, Italy \and
Consiglio Nazionale delle Ricerche - Istituto di Scienza e Tecnologie dell'Informazione \lq A.~Faedo\rq, CNR, Pisa, Italy}

\authorrunning{Belmonte et al.}

\maketitle
\newcommand{\computer}{a desktop computer equipped with an Intel Core I7 7700 processor (with 8 cores) and 16GB of RAM\xspace}

\begin{abstract}
Spatial and spatio-temporal model checking techniques have a wide range of application domains, among which large scale distributed systems and signal and image analysis. In the latter domain, automatic and semi-automatic contouring in Medical Imaging has shown to be a very promising and versatile application that can greatly facilitate the work of professionals in this domain, while supporting \emph{explainability}, easy \emph{replicability} and \emph{exchange} of medical image analysis methods. In recent work we have applied this model-checking technique to the (3D) contouring of tumours and related oedema in magnetic resonance images of the brain. In the current work we address the contouring of (2D) images of nevi. One of the challenges of treating nevi images is their considerable inhomogeneity in shape, colour, texture and size.  To deal with this challenge we use a {\em texture similarity}  operator, in combination with spatial logic operators.  We apply our technique on images of a large public database and compare the results with associated ground truth segmentation provided by domain experts.  
%
\end{abstract}






\section{Introduction and Related Work}
\label{sec:intro}

A nevus is a visible, usually small and benign, circumscribed lesion of the skin. Unfortunately, in some cases these are hard to distinguish from their malignant counterpart known as Melanocytic nevus. Melanoma is a very serious form of skin cancer. It may be lethal if the disease is not recognised in a very early stage. In Europe alone melanoma causes over 20,000 deaths each year~\cite{FMVBG2012}. One of the difficulties is that reliable early detection requires highly trained specialists but in many countries there is only a limited number of such specialists available. It is therefore no surprise that there is much interest in automated systems that can help recognising the disease reliably and at an early stage so that more lives could be saved and the number of unnecessary biopsies can be reduced~\cite{CNPGHHS2017}.

The most popular and well-performing automated techniques for the diagnosis of melanoma at the moment rely on deep learning~\cite{CNPGHHS2017}. In this paper we take a different approach based on recently developed {\em spatial} model checking techniques, in particular for the contouring or segmentation of nevi, which is one of the sub-tasks involved in the diagnosis of melanoma. In our previous work on (semi-) automatic contouring using spatial model checking techniques for the contouring of various kinds of brain tissues and brain tumours~\cite{Be+17,BCLM19,BelmonteCLM19,Ba+20} 
we have shown that such techniques can reach a segmentation quality that is competitive with
state-of-the-art techniques, while supporting explainability, easy replicability and exchange of medical image analysis methods. The segmentation of nevi poses additional challenges because dermoscopic images of nevi tend to be very dis-homogeneous in size, colour, contrast, location and kind of nevus/lesion and the presence of additional objects such as coloured patches, hairs, shadows and other optical effects.

Spatial (and spatio-temporal)  model checkers use high-level, often domain oriented, specifications written in a logical language to describe spatial properties in order to automatically and efficiently identify spatial patterns and structures of interest. The origins of spatial logic can be traced back to the forties of the previous century when McKinsey and Tarski recognised the possibility of reasoning on space using topology as a mathematical framework for the interpretation of modal logic (see~\cite{HBSL} for a thorough introduction). In recent work by Ciancia et al.~\cite{CLLM14,CLLM16} {\em Closure Spaces}, a generalisation of topological spaces, were used as underlying model for discrete spatial logic inspired by recent work by Galton~\cite{Gal99,Gal03,RLG13,Gal14}. This resulted in the definition of the Spatial Logic for Closure Spaces (SLCS), and a related model checking algorithm. Furthermore, in~\cite{CGLLM15}, a spatio-temporal logic, combining Computation Tree Logic with the spatial operators of SLCS was introduced. Spatial and spatio-temporal model checking have recently been applied in a variety of domains, ranging from Collective Adaptive Systems~\cite{CGLLM14,CLMP15,Ci+16a} to signals~\cite{Ne+18} and medical images~\cite{Ba+20,BCLM18arxiv,Be+17,BelmonteCLM19,BCLM19}.

Several proposals of use of computational methods for 
the analysis of medical images are available in the literature.
\emph{Computer-Aided Diagnosis} (CAD) aims at the
classification of areas in images, based on the presence of signs of
specific diseases \cite{Doi2007}. \emph{Image Segmentation}~\cite{Gordillo2013}
is focused on the identification of areas that have specific features or perform 
specific functions. \emph{Automatic contouring} of Organs at Risk or target
volumes ~\cite{brock2014image} is specifically devoted to supporting radiotherapy applications.
Finally, specific  \emph{indicators} can be computed from the acquired images that 
can enable early diagnosis---or the understanding of microscopic characteristics of specific
diseases---or can help in the identification of prognostic factors to predict a treatment
output~\cite{Chetelat2003,Toosy2003}. 
In~\cite{Gr+09} spiral electric waves---a precursor to atrial and ventricular fibrillation---are
detected and specified using a spatial logic and model-checking tools. The 
formulas of the logic are learned from the spatial patterns under investigation 
and the onset of spiral waves is detected using bounded model checking.
In~\cite{SPATEL} a logic called Spatial-Temporal Logic (SpaTeL) is defined that is a unification of signal temporal logic (STL) and tree spatial superposition logic (TSSL). The logic can
be used for describing high-level spatial patterns that change over time.

In our previous work on the use of spatial model-checking for the analysis of medical images mentioned earlier, we focused on image segmentation, in particular for the identification of glioblastomas---which are the most common malignant intracranial tumours---but also of regions of interest in healthy organs~\cite{BelmonteCLM19}.

In this paper, we apply spatial model-checking  to images of nevi from a public dataset released by the International Skin Imaging Collaboration (ISIC) for the 2016\footnote{We focus on the 2016 Challenge, that is the first of a series; We leave the subsequent challenges for future work.} International Symposium on Biomedical Imaging (ISBI 2016) challenge titled `` Skin Lesion Analysis toward Melanoma Detection''~\cite{CNPGHHS2017}. This dataset contains 900 annotated dermoscopic images, obtained by specialised high-resolution imaging of the skin that reduces skin surface reflectance. Among this set are 173 images of melanomas. Each image in the dataset has been segmented manually by experts and their segmentation result is available as ground truth images, which makes comparison with results of other state-of-the-art segmentation techniques applied to the same dataset possible. The original challenge consisted of three parts:  Lesion Segmentation, lesion Dermoscopic Feature Extraction, and Lesion Classification. In the present work we focus on lesion segmentation.


The outline of the paper is as follows. Section~\ref{sec:SpatialLogicFramework} provides some background on spatial model checking, the spatial model checker \imgqool  and in particular its texture similarity operator. Section~\ref{sec:segmentation} presents the spatial logic specification for the segmentation of nevi and Section~\ref{sec:results} presents the model checking results on (subsets) of the ISIC 2016 training dataset. Finally, Section~\ref{sec:conclusions} presents the conclusions of this work.

\def\calI{{\cal I}}
\newcommand\lrcs{\stackrel{\rightarrow}{\rho}}
\newcommand\lrcd{\stackrel{\leftarrow}{\rho}}
\newcommand\interior{\calI}
\newcommand{\lssimsym}{\triangle\!\!\!{\scriptstyle \triangle}}

\section{Background on Spatial Model Checking}
\label{sec:SpatialLogicFramework}
\SLCSMI (\emph{Image Query Language}), first proposed in~\cite{Ba+20,BCLM19}, is a spatial logical language developed for the analysis of medical images. It is based on \SLCS (\emph{Spatial Logic for Closure Spaces}) ~\cite{CLLM14,CLLM16}. \SLCSMI is also the input language for the spatial model checker \imgqool  presented in~\cite{BCLM19}. In this section we first recall the definition of the logical kernel of \SLCSMI and the underlying basic notions and then we show its extension supported by the tool. We refer to our earlier work for further details on theoretical aspects and the spatial model checking algorithms~\cite{CLLM14,CLLM16,Ba+20,BCLM19}.

\subsection{The logical kernel of \SLCSMI}
\SLCS is interpreted over {\em closure spaces}. A closure space---CS for short---is a pair $(X,\closure)$ where $X$ is a set (of points) and $\closure:2^X \to 2^X$ is a function satisfying the following three axioms: (i) $\closure(\emptyset)=\emptyset$; (ii)
$Y \subseteq \closure(Y)$ for all  $Y \subseteq X$; (iii)
$\closure(Y_1 \cup Y_2) = \closure(Y_1) \cup \closure(Y_2)$ for all $Y_1,Y_2\subseteq X$. The {\em interior} of a set $Y\subseteq X$ is obtained by duality, i.e. $\interior(Y)=\overline{\closure(\overline{Y})}$
where $\overline{Y}=X\setminus Y$ is the complement of $Y$. Given any 
relation $\arel \subseteq X \times X$, $(X,\closure_{\arel})$, with $\closure_{\arel}(Y) = Y\cup \ZET{x}{\exists y \in Y. y \, \arel \,x}$,  is a CS. In particular, a digital image can be modeled as a finite CS where $X$ is the set of voxels and $\arel$ their (reflexive and symmetric) {\em adjacency relation}\footnote{All the theory and related model checkers work both for 2D and 3D even though we use only 2D in the current work. Similarly, in the current work we use the word `voxel' both for 3D `pixels' and for 2D pixels.}. 

$(\nats,\closure_{succ})$ is the CS of  the natural numbers $\nats$
with  the binary successor relation $succ = \ZET{(m,n) \in \nats^2}{n=m+1}$. A (discrete) {\em path} over
$(X,\closure)$ is a continuous function\footnote{A {\em continuous} function from CS $(X_1,\closure_1)$ to CS $(X_2,\closure_2)$ is a function $f: X_1 \to X_2$ such that $f(\closure_1(Y)) \subseteq \closure_2(f(Y))$ for all $Y \subseteq X_1$.}  from $(\nats,\closure_{succ})$ to $(X,\closure)$.

It is often convenient to equip the elements of $X$ with {\em attributes} in a given set $A$ over a given set of values $V$; an {\em attributed} CS is a structure $((X,\closure),\aeval)$ where 
$(X,\closure)$ is a CS and $\aeval: A \times X \to V$, is the attribute evaluation function, such that $\aeval(a,x)$  maps attribute (named) $a$ of point $x$ to its value in $V$. 
For instance, if $x$ is a voxel, then $\aeval(red,x)$ may represent the intensity of red of $x$, and similarly for $\aeval(green,x)$ and $\aeval(blue,x)$. Attribute values can be used in expressions $\alpha$ over $V$; consequently function $\aeval$ is assumed lifted to such expressions in the standard way. 

In this paper we will use {\em distance} CS, i.e. structures $((X,\closure), d)$ where $d: X \times X \to \reals_{\geq 0} \cup \SET{\infty}$ is a {\em distance function}\footnote{Several distance functions  are defined in the literature; the specific distance to be used depends on the application. The interested reader is referred to~\cite{Ba+20}. In this work we use the Manhattan distance where 1 voxel is the unit distance.}, i.e. it satisfies $d(x,y)=0$ if and only if $x=y$; $d$ is lifted to sets in the usual way: $d(x,\emptyset)=\infty$ and for $\emptyset \subset Y \subseteq X$
$d(x,Y)=\inf\ZET{d(x,y)}{y \in Y}$.

\SLCSMI is interpreted over {\em attributed distance closure models}, i.e. structures
$((X,\closure),d,\aeval, \peval)$ where $(X,\closure)$ is a CS,
$d$ and $\aeval$ are the distance and the attribute evaluation functions, respectively, and
$\peval: \props \to 2^X$ is a valuation which maps the  {\em atomic predicates} of a given set $\props$ to the points satisfying them. 
In the sequel we recall the formal definition of the logical kernel of \SLCSMI:

\begin{definition}\label{def:SLCSMI}
For given set $\props$ of {\em atomic predicates} $p$, and interval $I$ of $\reals$, the syntax of \SLCSMI is the following:
$$
\form  ::=   p  \msep  \lneg \, \form  \msep  \form_1 \, \land \, \form_2  \msep  
\lrcs \form_1[\form_2] \msep \lrcd \form_1[\form_2] \msep \dist{I} \form. 
$$
{\em Defined} predicates are elements $p$ of $\props$ for which a {\em defining equation}
$p:=\alpha$ is given, where $\alpha$ is an expression.

{\em Satisfaction} $\model, x \models \form$  of a formula $\form$ at point $x \in X$ in model
$\model = (((X,\closure),d), \aeval, \peval)$ is defined recursively on the structure of formulas,
where $\sem{\form}^{\model}$ is the set $\ZET{x\in X}{\model, x\models \form}$ of points satisfying $\form$ in $\model$:
\[
\begin{array}{r c l c l c l L}
\model,x & \models  & p \in P & \Leftrightarrow & x  \in \peval(p)\\
\model,x & \models  & \lneg \,\form & \Leftrightarrow & \model,x  \models \form \mbox{ does not hold}\\
\model,x & \models  & \form_1\, \land \,\form_2 & \Leftrightarrow &
\model,x  \models \form_1 \mbox{ and } \model,x  \models \form_2\\
\model,x & \models  & \lrcs\form_1[\form_2 ]& \Leftrightarrow & \mbox{there is path } \pi 
\mbox{ and index } \ell 
\mbox{ s.t. }
\pi(0)=x \mbox{ and 
 } 
\model, \pi(\ell) \models  \form_1 
\\
& &  & & 
\mbox{and for all indexes } j: 0 < j  < \ell \mbox{ implies } \model, \pi(j) \models \form_2\\
\model,x & \models  & \lrcd\form_1[\form_2 ]& \Leftrightarrow & \mbox{there is path } \pi 
\mbox{ and index } \ell 
\mbox{ s.t. }
\pi(\ell)=x \mbox{ and 
 } 
\model, \pi(0) \models  \form_1 
\\
& &  & & 
\mbox{and for all indexes } j: 0 < j  < \ell \mbox{ implies } \model, \pi(j) \models \form_2\\
\model, x & \models & \dist{I} \, \form & \Leftrightarrow &
d(x, \sem{\form}^{\model}) \in I
\end{array}
\]
Whenever $p$ is a {\em defined} predicate with defining equation $p:=\alpha$, we extend the satisfaction relation by letting $x\in \peval(p)$
if and only if  $\aeval(x,\alpha)$ is true. 
\ed
\end{definition}

Classical derived operators are defined as usual: 
$\lfalse \equiv p \land \lneg p$,
$\ltrue \equiv  \lneg \lfalse$,
$\form_1 \lor \form_2 \equiv \lneg(\lneg \form_1 \land \lneg \form_2)$ etc. In addition, we have the following more specific derived operators:
$$
\begin{array}{l c l c l c l}
\lnear \form & \equiv & \lrcd \form [\lfalse] \\ 
\form_1\, \lsurr \, \form_2 &\equiv& \form_1 \land \lneg \lrcs \lneg (\form_1 \lor \form_2)[\lneg \form_2]\\
\mathit{touch }(\form_1 , \form_2) & \equiv&   \form_1 \land \lrcs \form_2[\form_1]\\
\mathit{grow } (\form_1 , \form_2)  &\equiv & \form_1 \lor \mathit{touch}( \form_2 , \form_1)\\
\mathit{smoothen }(r, \form_1) &\equiv & \dist{< r} (\dist{\geq r} \lnot\form_1).
\end{array}
$$
Intuitively, a point $x$ satisfies $\lnear \form$ if it is {\em near} $\form$, i.e. if it can be reached {\em in one step} from a point laying in $\sem{\form}{}$; it is easy to see that $\model,x \models \lnear \form$ if and only if $x \in \closure(\sem{\form}^{\model})$. A point $x$ satisfies
$\form_1\, \lsurr \, \form_2$ if it lays in an area, where all points satisfy $\form_1$, that is {\em surrounded} by points satisfying $\form_2$, i.e. it is impossible to find a path starting from $x$ that can reach a point satisfying neither $\form_1$ nor $\form_2$, without first passing through a point satisfying $\form_2$\footnote{Note that in~\cite{CLLM14,CLLM16,Ba+20,BCLM19} $\lnear$ and $\lsurr$ (denoted by ${\cal U}$ in~\cite{CLLM14}) have been presented as basic operators while reachability operators have been defined as derived from the formers.}.
The meaning of $\mathit{touch } (\form_1 , \form_2)$ should be clear.
A point satisfies $\mathit{grow } (\form_1 , \form_2)$ if it satisfies $\form_1$ or it lays in a path
of points all satisfying $\form_2$ and leading to a point satisfying $\form_1$.
A formula $\mathit{smoothen }(r,\form_1)$ is satisfied by points that are at a distance of less than $r$ from a point that is at least at distance $r$ from points that do not satisfy $\form_1$. This operator 
works as a filter; only contiguous areas satisfying $\form_1$ that have a minimal diameter of at least $2r$ are preserved; these are also smoothened if they have an irregular shape (e.g. protrusions with a width that is less than the indicated distance).

We close this section with the description of an additional logical operator of \SLCSMI, namely
the {\em Texture Analysis} operator $\lssimsym$ introduced in~\cite{BCLM19}. Texture Analysis (TA) is an approach used for finding patterns in (medical) images.  
The approach has proved promising in a large number of applications in the field of medical imaging
\cite{Kassner2010,Lopes2011,Castellano2004,Davnall2012}; in particular
it has been used in \emph{Computer Aided Diagnosis}
\cite{Woods2007,Han2014,Heinonen2009} and for classification or
segmentation of tissues or organs
\cite{Chen1989,Sharma2008,RodriguezGutierrez2013}.
The \SLCSMI TA operator $\lssimsym$ uses \emph{first order} statistical methods\footnote{First order statistical methods are statistics based on the probability distribution function of the intensity values of the pixels of parts, or the whole, of an image.} and differs from those in the classical setting,  e.g.~\cite{Srinivasan2008,Tijms2011}, where the various moments (\emph{mean}, \emph{variance} etc.) of distributions of the two pictures to be compared are analysed. In \SLCSMI, instead, the 
statistical distributions---actually the histograms---of the two pictures are compared directly, 
using, as similarity measure, their 
\emph{cross-correlation} (also called \emph{Pearson's correlation coefficient}).

\begin{figure}
\centerline{\includegraphics[width=0.35\textwidth]{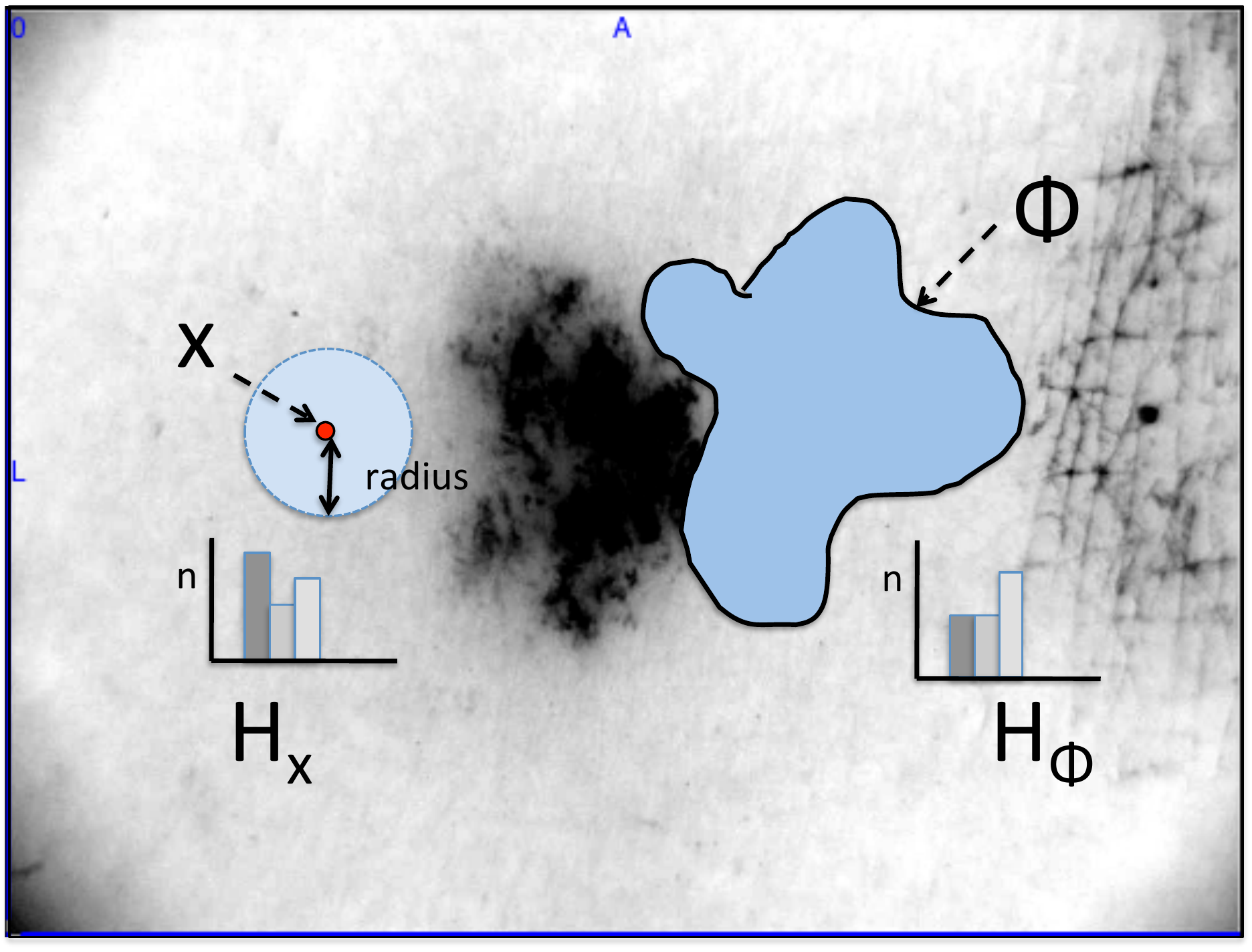}}
\caption{\label{fig:TAsketch} Illustration of the \SLCSMI TA operator $\lssimsym$}
\end{figure}

The intuitive semantics of $\lssimsym$ is presented schematically in Figure~\ref{fig:TAsketch}. Suppose we have an \SLCSMI formula $\form$ that specifies a {\em sample area  of interest} $\sem{\form}^{\model}$ in a given image; suppose, furthermore, that the feature that makes the points in 
$\sem{\form}^{\model}$ `interesting' is represented by the values of a certain attribute $a$ of theirs, for instance voxel luminosity.
A common representation of such values is their {\em histogram} $H_{\form}$, where the range of values is
split into adjacent intervals of equal width---called {\em bins}---and for each bin, say $k$, 
$H_{\form}(k)$ is the total number of points that have a value of $a$ falling in $k$.
A point $x$ is considered {\em similar} to the sample area if the histogram $H_x$ (of an attribute of interest) of the points laying in a small area around $x$ {\em correlates} with $H_{\form}$. Of course, the two histograms must have the same number of bins. 

More precisely, with reference to model $\model = ((X,\closure), \aeval, \peval)$, the 
histogram $\mkhis(a,Y,m,M,k)$ of the distribution of the values of attribute $a$ of the points in $Y\subseteq X$, in the interval $[m,M]$ with step size $\Delta$ and $k$ bins, is the
function $\mkhis : A \times 2^X \times \reals \times \reals \times \nats \to (\nats \to \nats)$
such that, for all values $m,M \in \reals$, with $m<M$, and $k\in \nats \setminus \SET{0}$, and 
$i\in \SET{1,\ldots, k}$,
$
\mkhis(a,Y,m,M,k)(i) =
\left|\ZET{y\in Y}{(i-1) \Delta \leq \aeval(y,a) - m < i \Delta}\right|
$,
where $\Delta=\frac{M-m}{k}$. The {\em mean}  $\mean{h}$ of histogram $h$ 
is $\frac{1}{k}\sum_{i=1}^k h(i)$. 
Given histograms $h_1, h_2: \SET{1,\ldots, k} \to \nats$, their {\em cross correlation} $\cc(h_1,h_2)$ is given by
$
\cc(h_1,h_2) = 
\frac
{\sum_{i=1}^k\left( h_1(i) - \mean{h_1} \right) \left(h_2(i) - \mean{h_2} \right)}
{
\sqrt{ 
\sum_{i=1}^k \left(h_1(i) - \mean{h_1} \right)^2
}
\sqrt{
\sum_{i=1}^k \left(h_2(i) - \mean{h_2} \right)^2
}
}
$.

The value of $\cc{}{}$ is \emph{normalised} so that
$-1\le \cc(h_1,h_2) \le 1$; $\cc(h_1,h_2)=1$ indicates that $h_1$ and $h_2$
are \emph{perfectly correlated} (that is, $h_1 = \alpha h_2+\beta$, with $\alpha>0$); $\cc(h_1,h_2) =-1$
indicates \emph{perfect anti-correlation} (that is, $h_1=\alpha h_2+\beta$, with $\alpha<0$). On the other
hand, $\cc(h_1,h_2) = 0$ indicates no correlation\footnote{Note that normalisation
makes the value of $\cc{}{}$ undefined for constant histograms, having
therefore standard deviation of $0$; in terms of statistics, a
variable with such standard deviation is only (perfectly)
correlated to itself. This special case is handled by letting
$\cc(h_1,h_2)=1$ when both histograms are constant, and $\cc(h_1,h_2)=0$ when only one
of the $h_1$ or $h_2$ is constant.}. 

We now have all the ingredients for completing the definition of the logical kernel of  \SLCSMI by extending the 
syntax given in Def.~\ref{def:SLCSMI} with $\lssim{\bowtie c}{m}{M}{k}{r}{a}{b}$, and adding 
the following clause to the definition of the satisfaction relation, where $S(x,r)=\ZET{y\in X}{d(x,y) \leq r}$ is the {\em sphere} of radius $r$ centred in $x$, $h_a(i)=\mkhis(a,S(x,r),m,M,k)(i)$, 
$h_b(i)=\mkhis(b,\sem{\form}^{\model},m,M,k)(i)$, and $\bowtie \, \in \, \SET{=,<,>,\leq,\geq}$:
$$
\model, x \, \models \,
\lssim{\bowtie c}{m}{M}{k}{r}{a}{b} \form \Leftrightarrow
\cc(h_a,h_b) \bowtie c.
$$

So $\lssim{\bowtie c}{m}{M}{k}{r}{a}{b} \form$ compares the region of the space
 constituted by the sphere of radius $r$ centred in $x$ against the 
region characterised by $\form$. The comparison is  based on the cross correlation
of the histograms of the values of the chosen attributes of (the points of) the two regions, namely attribute $a$ for the points around $x$ and attribute $b$ for the points that satisfy $\form$. Of course, these attributes may also be chosen to be the same.
Both histograms share the same domain ($[m,M]$) and the same (number of) bins ($\SET{1,\ldots, k}$). 
%

\subsection{\imgqool}

\imgqool\footnote{\imgqool\ is available at {\sf https://github.com/vincenzoml/VoxLogicA}. In the present paper we used version 0.5.99.1-experimental.} is a model-checker for \SLCSMI that is specialised for digital image analysis. It is
a {\em global} spatial model-checker in the sense that, given a model $\model$ (i.e. a digital image)
and a formula $\form$, it computes the set $\sem{\form}^{\model}$ of all voxels in the image that satisfy $\form$. Such a set can be, and usually is, represented by a boolean image---i.e. a closure model of the same dimension and size of $\model$, where each point is assigned the value
{\em true} if the corresponding voxel in $\model$ satisfies $\form$, and {\em false} otherwise.

Actually, this feature is pushed forward in \imgqool so that one can
obtain a (result) {\em quantitative} image where each point has a numerical value that is, for instance, the 
cross-correlation score computed for the verification of a $\lssimsym$-formula on the corresponding voxel of $\model$. This is precisely what is done in the following example:\\

\noindent
{\sf let scores = crossCorrelation(5,inty,inty,sample,min(inty),max(inty),15)}\\

\noindent
where {\sf sample} is a formula characterising the sample portion of the image at hand and
every point of {\sf scores} will be associated with the score of the correlation between
the intensity ({\sf inty}) histogram of the sphere of radius $5$ centred in the corresponding
voxel of the image and intensity histogram of the {\sf sample} area in the image, both histograms having 15 bins.

Functions and predicates can be defined in \imgqool in the usual way. For instance\\

\noindent
{\sf let strongCorr(r,a,b,F,m,M,k,c) = crossCorrelation(r,a,b,F,m,M,k) $>$ c}\\

\noindent
is the \imgqool equivalent of $\lssim{> c}{m}{M}{k}{r}{a}{b} F$ so that\\

\noindent
{\sf let interesting = strongCorr(5,inty,inty,sample,min(inty),max(inty),15,9.8)}\\

\noindent
returns in  {\sf  interesting} a boolean image where the value {\em true} is associated to
each voxel corresponding to a point in the current image which is the centre of a sphere 
of radius 5 the intensity of which has a high---higher than $9.8$---correlation with the {\sf  sample} 
portion of the current image, and {\em false} to any other point.

The following additional commands are available in \imgqool (more details can be found in~\cite{BCLM19}):
\begin{itemize}
\item
{\sf load x = ``s"} loads an image from file {\sf ``s"} and binds it to {\sf x} for subsequent usage;
\item
{\sf save ``s" e} stores the image resulting from evaluation of expression {\sf e} to file {\sf ``s"};
\item
{\sf print ``s" e} prints to the log the string {\sf s} followed by the numeric, or boolean,
result of computing {\sf e};
\item 
{\sf import ``s"} imports a library of declarations from file {\sf ``s"}; 
\end{itemize}




\section{Segmentation of Nevus with \imgqool}
\label{sec:segmentation}

\begin{figure}
\def\gbsz{1.9cm}
\centering
\subfloat[][]
{
\includegraphics[height=\gbsz]{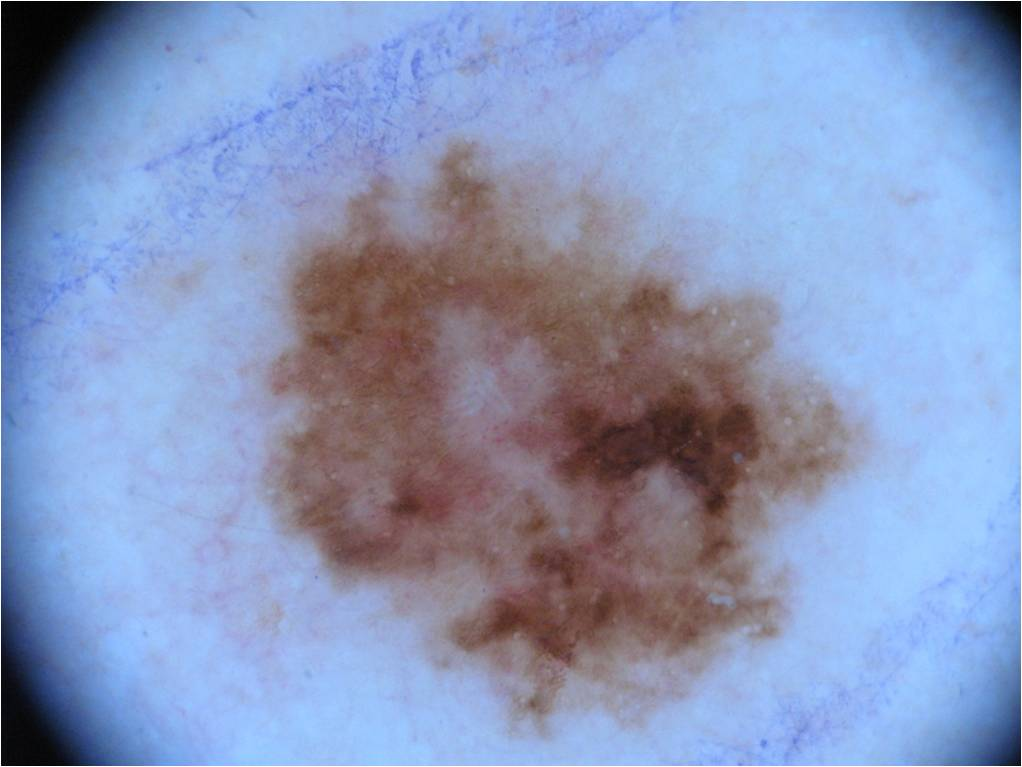}
\label{subfig:02}
}
\centering
\subfloat[][]
{
\includegraphics[height=\gbsz]{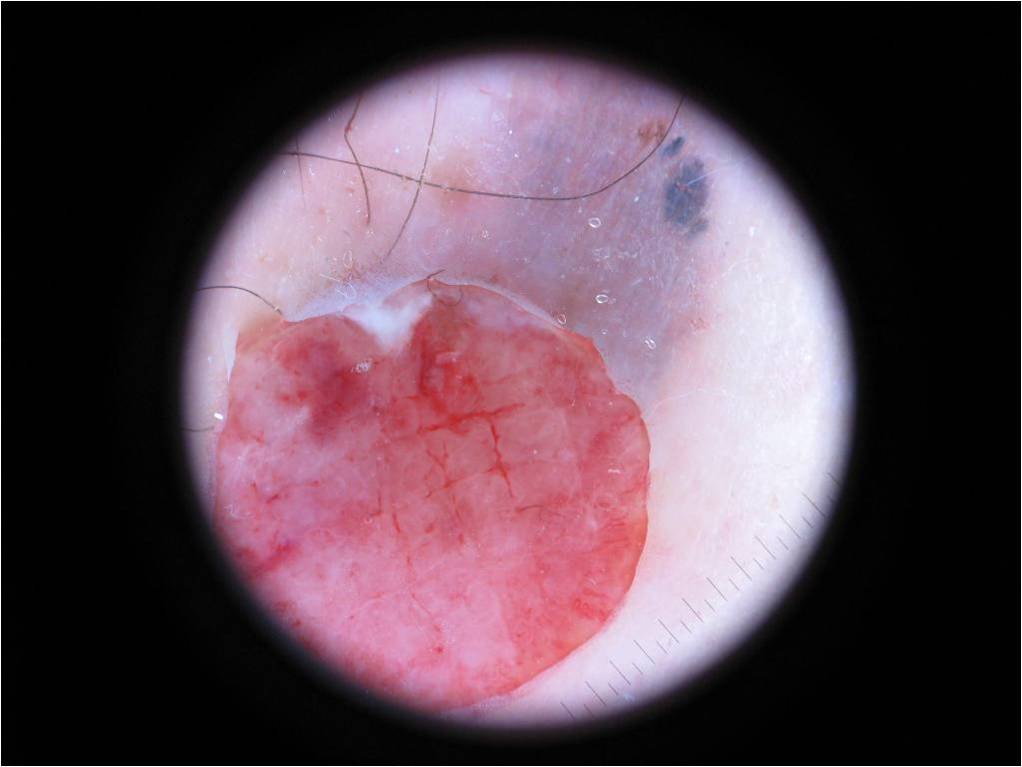}
\label{subfig:04}
}
\centering
\subfloat[][]
{
\includegraphics[height=\gbsz]{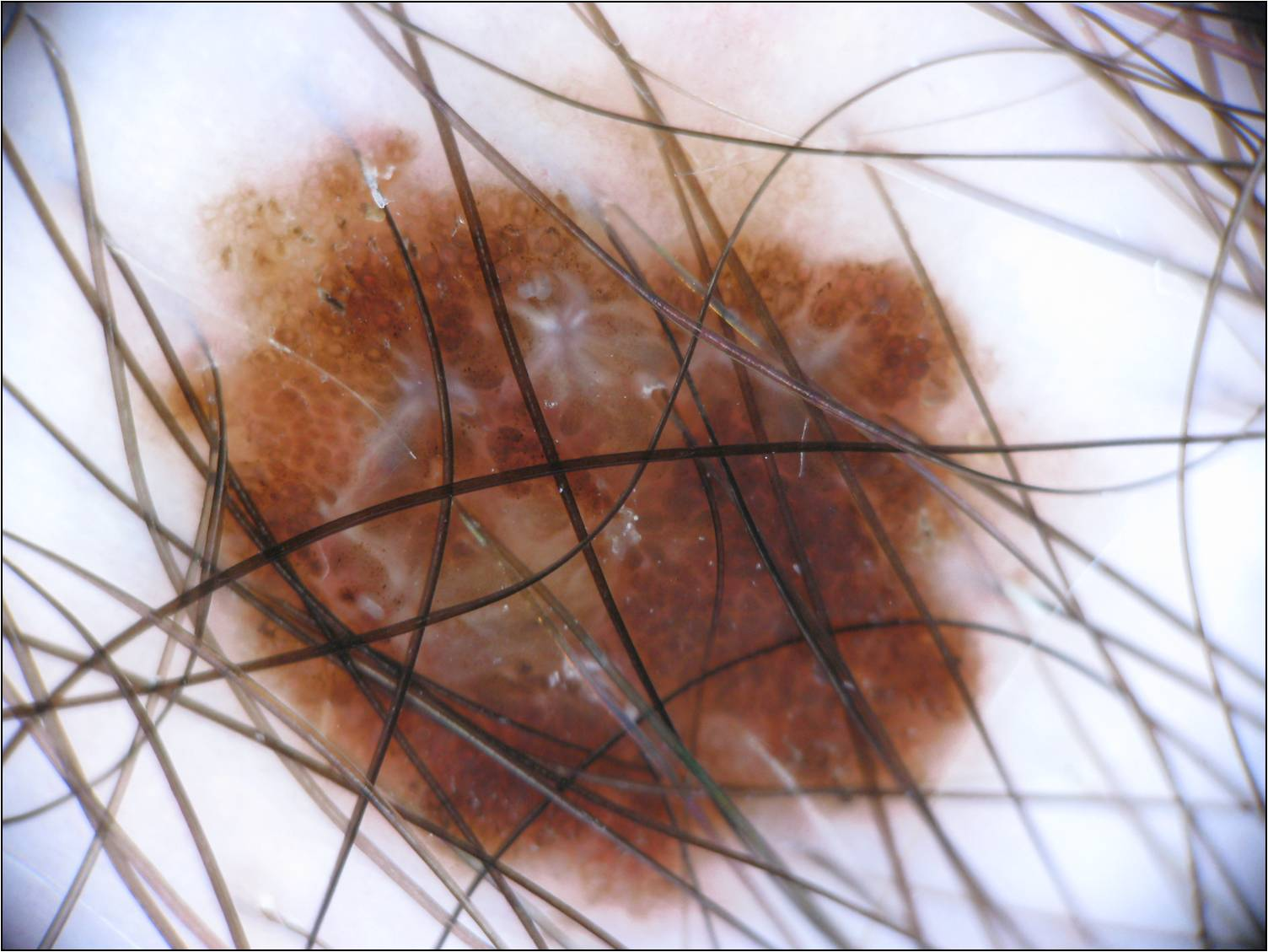}
\label{subfig:43}
}
\centering
\subfloat[][]
{
\includegraphics[height=\gbsz]{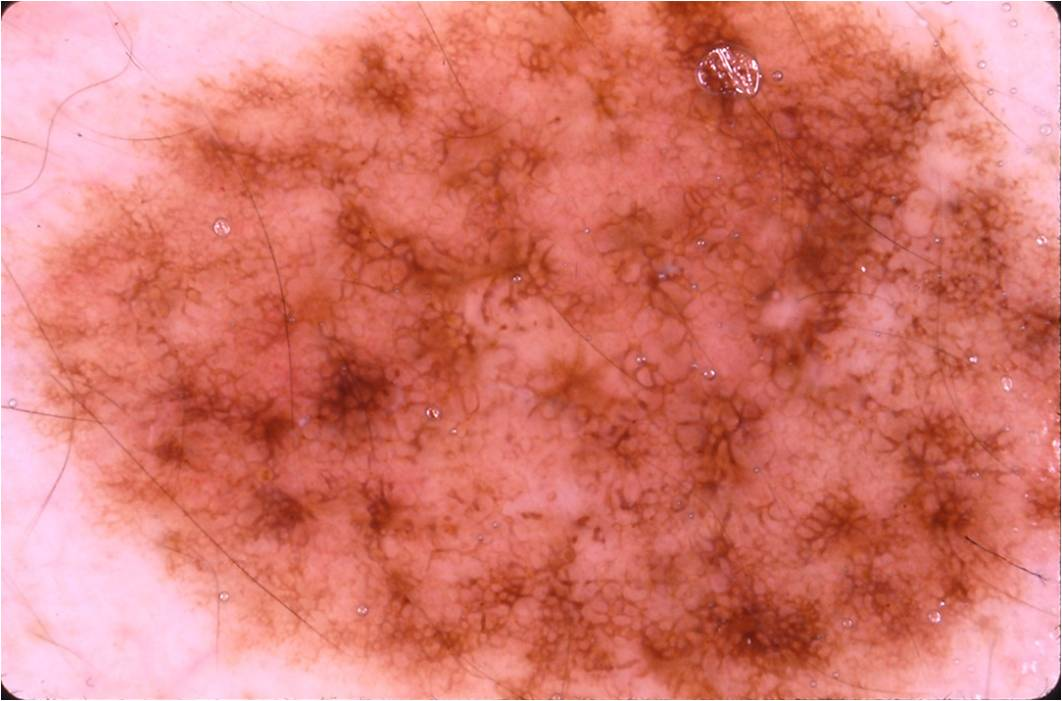}
\label{subfig:207}
}\\
\centering
\subfloat[][]
{
\includegraphics[height=\gbsz]{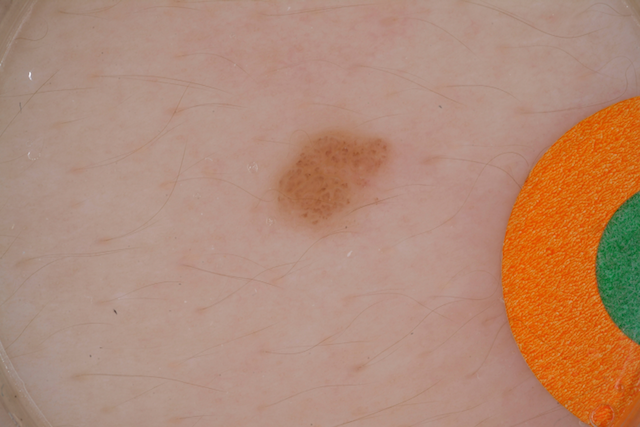}
\label{subfig:1191}
}
\centering
\subfloat[][]
{
\includegraphics[height=\gbsz]{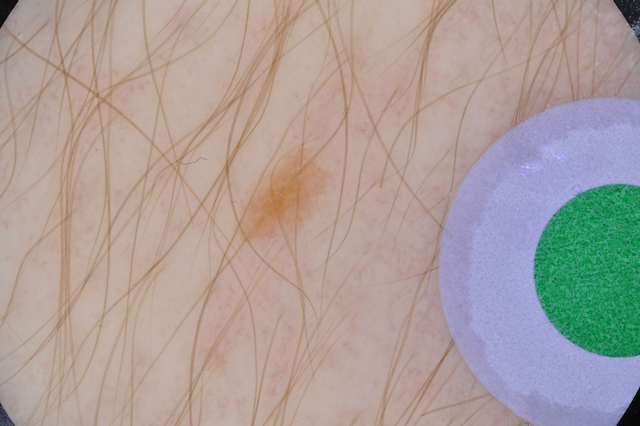}
\label{subfig:5787}
}
\caption{Example images from the ISIC 2016 dataset illustrating the dishomogeneity of nevi. They also differ greatly in resolution, e.g. the size of (a) is 486 KB and that of (f) 11,3 MB (compared in .png format).
}
\label{fig:nevi}
\end{figure}

As mentioned briefly in the introduction, one of the complications of the segmentation of nevi is their great variability in appearance and the dishomogeneity of the dermoscopic images themselves. Nevi may show very different colour ranges, also within the same nevus, have different sizes, can be more or less regular, appear on more or less regular skin where hairs or sebaceous follicles may be present as well. Furthermore, the images themselves also show quite a variety and may be of different size, showing black corners, rings, or shadows due to the lenses used, showing more or less contrast and intensity or the presence of patches near the nevus. The images in Fig.~\ref{fig:nevi} show a few examples of this dishomogeneity as encountered in the ISIC 2016 training dataset\footnote{Datasets can be found at {\sf https://challenge.isic-archive.com/data}}.

In the present paper we therefore make the following simplifying assumptions on the nevus images:
\begin{enumerate}
\item The nevus is surrounded by skin texture and is not touching or overlapping with the (black) border of the image;
\item The nevus is connected and does not consist of multiple parts that are separated by skin texture;
\item The skin texture type surrounding the nevus is of type I or II (fair skin colour) or at most type III (uniform light tan)~\cite{Fit88}.
\end{enumerate}

In future work we will study to what extent these assumptions can be relaxed.

\subsection{Nevus Segmentation using Texture Analysis}

\begin{figure}
\def\gbsz{1.9cm}
\centering
\subfloat[][]
{
\includegraphics[height=\gbsz]{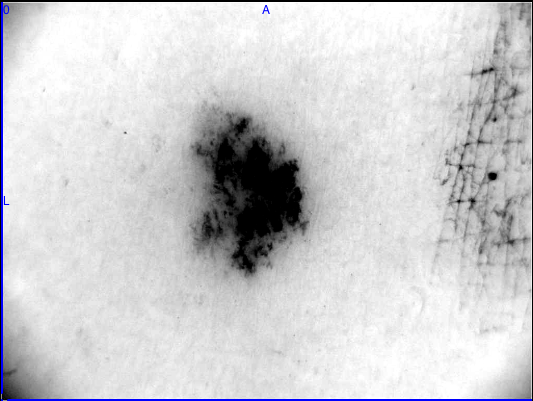}
\label{subfig:orig}
}
\centering
\subfloat[][]
{
\includegraphics[height=\gbsz]{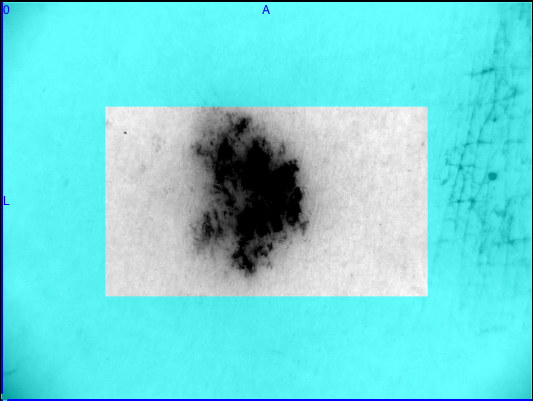}
\label{subfig:bgSample}
}
\centering
\subfloat[][]
{
\includegraphics[height=\gbsz]{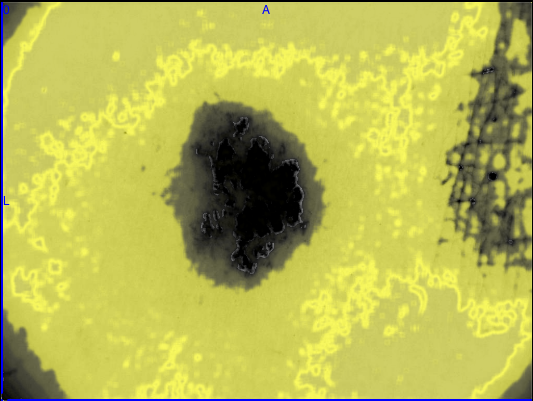}
\label{subfig:bgScore}
}
\centering
\subfloat[][]
{
\includegraphics[height=\gbsz]{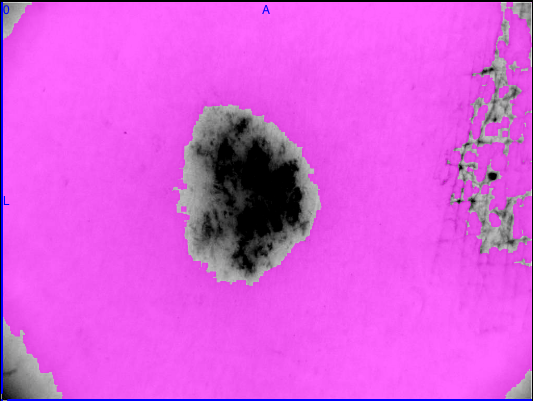}
\label{subfig:bgSim}
}\\
\centering
\subfloat[][]
{
\includegraphics[height=\gbsz]{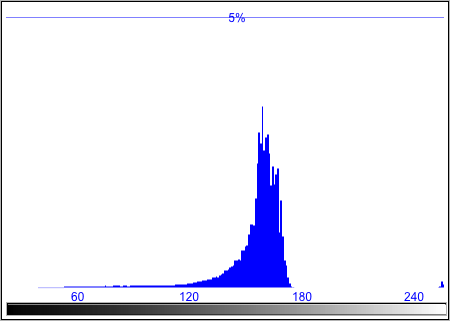}
\label{subfig:origHist}
}
\centering
\subfloat[][]
{
\includegraphics[height=\gbsz]{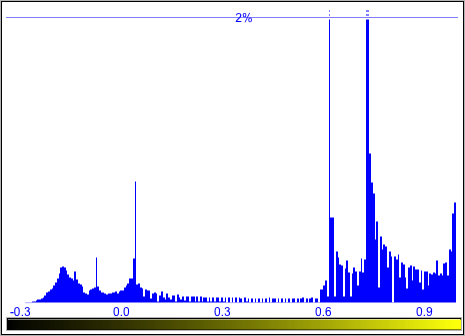}
\label{subfig:bgSimHis}
}
\caption{Illustration texture analysis operator \imgqool. Fig.~\ref{subfig:orig} Original image of nevus (image ISIC-0000010 of the 2016 Challenge) and surrounding skin and related histogram of voxel intensity in Fig.~\ref{subfig:origHist}. Sample $\Phi$ of background voxels shown in cyan (Fig.~\ref{subfig:bgSample}). Cross-correlation score for surrounding of each voxel w.r.t. histogram of background sample $\Phi$. The score is shown in yellow. The higher the score the higher the intensity of the yellow colour of voxels in Fig.~\ref{subfig:bgScore}.  The distribution of the cross-correlation scores is shown in Fig.~\ref{subfig:bgSimHis}. Voxels with a cross-correlation score of more than 0.05 are shown in pink in Fig.~\ref{subfig:bgSim}.
}
\label{fig:TA}
\end{figure}

Since there is very little one can take for granted in the images in the ISIC 2016 dataset, in the following 
we show some very coarse heuristics to get the process of segmentation started and illustrate the core steps of the segmentation approach. Some intermediate results are shown in  Figure~\ref{fig:TA}, illustrating the role of the texture analysis operator in this process on the image of a nevus shown in Fig.~\ref{subfig:orig} (Fig.~\ref{subfig:origHist} shows the associated intensity distribution). The main aim is to distinguish voxels that are part of the background (skin) from those that are likely part of the nevus. First we assume that our task is to find all voxels that are likely to be part of the background, so the healthy skin surrounding the nevus. We assume furthermore that at least part of the nevus is somewhere in the middle of the image so that we can take an area relatively close to the border as a sample of the background. Let $\form$ be the \SLCSMI formula that specifies such an area, shown  in Fig~\ref{subfig:bgSample}  as a semi-transparant overlay in cyan on the original image---later in the paper we will show $\form$ in detail. 
Let us, for now, also assume that we work with the intensity of the voxels rather than their colour or other attributes. As described in Section~\ref{sec:SpatialLogicFramework}, we construct the histogram $H_{\Phi}$ of the distribution of the intensity values of all the voxels that satisfy $\Phi$. Assume that $H_{\Phi}$ has $k$ bins, and a minimum and maximum value that correspond to the minimum and maximum pixel intensity in the whole image.

Next we construct the local histogram $H_x$ for each pixel $x$ in the image by taking the intensity of all the pixels that are present in a radius $rad$ around pixel $x$. This second histogram has the same number of bins and minimal and maximal values as those of histogram $H_{\Phi}$. We can now compute the similarity score for each point $x$ in the image by computing the Pearson's correlation coefficient of the histograms  $H_{\Phi}$ and $H_x$. Recall that this provides normalised values between -1 and +1. A value equal to 1 indicates perfect correlation between the histograms, a value equal to -1 indicates perfect anti-correlation. A score of value 0 indicates that there is no correlation between the histograms. The result for  Fig.~\ref{subfig:orig} is shown in Fig.~\ref{subfig:bgScore} as a semi-transparent yellow overlay where higher values of the score correspond to a brighter yellow hue. The associated histogram of these cross-correlation scores are shown in Fig.~\ref{subfig:bgSimHis}. Finally, in Fig.~\ref{subfig:bgSim}, those pixels with a cross-correlation score above 0.05 are shown as an overlay in pink.

Thus,  this particular use of the texture operator can provide a rather good first approximation of the area covered by the nevus. Clearly, it is not perfect yet, as also some other areas remain that are not identified as part of the background, whereas they should be. But these areas can in principle be identified by other means, such as their relative position with respect to the border of the image and other aspects that distinguish them from the nevus itself, as done in Specification~\ref{alg:seg}, shown later on, where we show in more detail how the above described procedure can be specified in  \SLCSMI. This specification uses a predicate, {\sf patch} and a derived operator {\sf relDist}. The former is a predicate specifying voxels that are part of a patch. The latter is a derived operator that defines the {\em relative distance} in an image, depending on its size. The definition of both the operators are provided and explained after Specification~\ref{alg:seg}.

Moreover, in Section~\ref{sec:results} we will use common similarity indexes to compare the quality of the segmentation. 
These indexes are defined directly in terms of \SLCSMI and shown in Specification~\ref{alg:indexes}.
They provide numeric support in the form of values of several commonly used similarity indexes that allow for an objective comparison with expert ground truth and with the performance of other state-of-the-art approaches. Note that the \imgqool procedure provided in the sequel does not require any particular preprocessing of the images as provided by the ISIC 2016 dataset, except for format conversion for the test data (from jpg format to png format) and colour conversion for the ground truth data (from grayscale to RGB). So we do not use explicit image pre-processing transformations that for example remove hairs or black corners or borders of the image or that increase contrast or normalise the size of the images.

\begin{algorithm}
    \small
    \tt
\SetAlgoLined
\caption{\label{alg:seg} Nevus segmentation specification}
\imgImport "stdlib.imgql"\\[0.3em]
load groundTruth = "\$INPUTDIR/\$NAME\_seg\_RGB.png"\\
load nevus = "\$INPUTDIR/\$NAME.png"\\
let nevusImgIntens = intensity(nevus)\\
let groundIntens = intensity(groundTruth)\\
%
%
%

let similarTo(a,rad) = crossCorrelation(rad,nevusImgIntens, nevusImgIntens,a,min(nevusImgIntens),max(nevusImgIntens),15) 

%
%
%
%
%
let almostBlack = nevusImgIntens <. 40.0 \\
let blackBorder = grow(distleq(relDist(5),border), almostBlack)\\
%
let bgSampleWidth =  relDist(200)\\
let bgSample = distleq(bgSampleWidth, blackBorder) \& (!blackBorder) \& (!patch) \\
%
let bgSimScore = similarTo(bgSample, relDist(5))\\ 
let bgSim = (bgSimScore >. 0.05) \& (! patch) \& (!blackBorder) \\
%
let preSeg = ((!border) S (bgSimScore <. 0.11)) \& (! patch) \& (!blackBorder)\\ 
let nevSeg = smoothen(maxvol(preSeg \& (! (distleq(relDist(50), blackBorder)))), relDist(3))\\
let nevSegSmooth = smoothen(maxvol(preSeg \& (! (distleq(relDist(50), blackBorder)))), relDist(10))\\
let nevSeg1 = maxvol(nevSeg \& nevSegSmooth \& !patch)\\
%
let nevSegV0 = nevSeg1 \\
\end{algorithm}

\begin{figure}
\def\gbsz{1.9cm}
\centering
\subfloat[][]
{
\includegraphics[height=\gbsz]{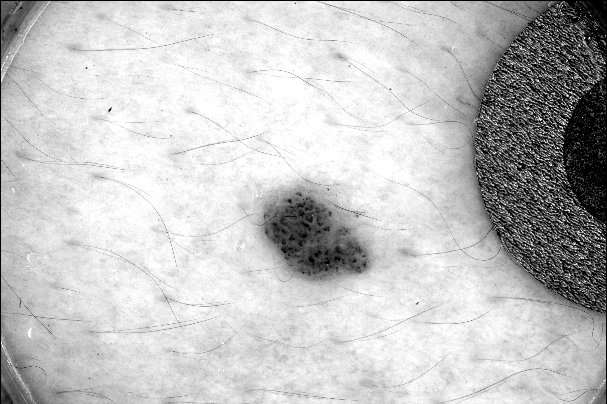}
\label{subfig:orig002}
}
\centering
\subfloat[][]
{
\includegraphics[height=\gbsz]{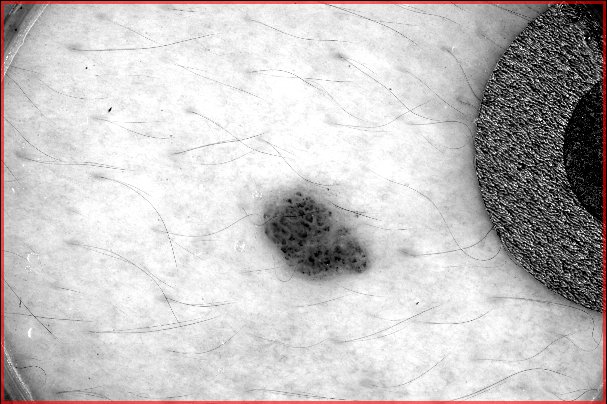}
\label{subfig:blackBorder002}
}
\centering
\subfloat[][]
{
\includegraphics[height=\gbsz]{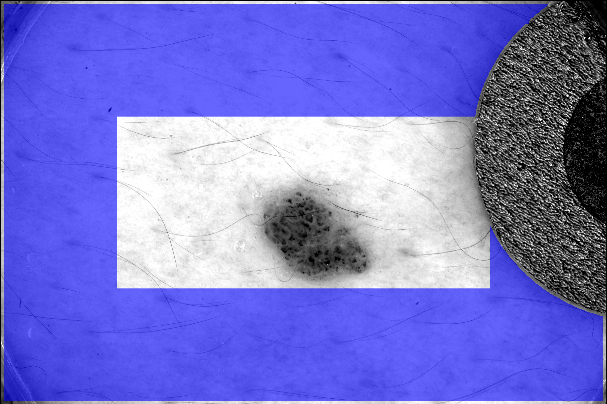}
\label{subfig:bgSample002}
}
\centering
\subfloat[][]
{
\includegraphics[height=\gbsz]{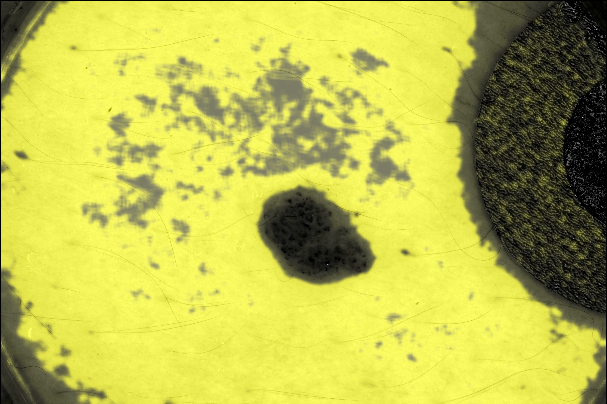}
\label{subfig:bgSimScore_002}
}\\
\centering
\subfloat[][]
{
\includegraphics[height=\gbsz]{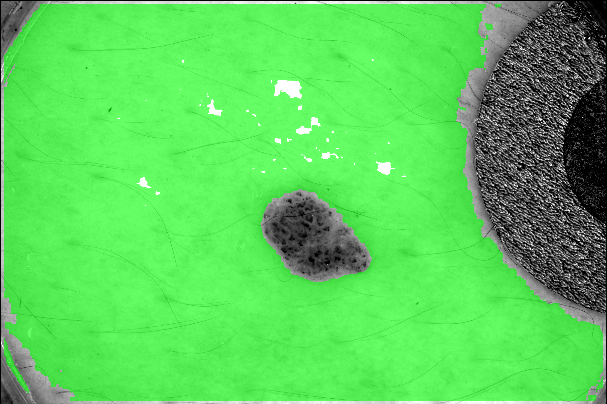}
\label{subfig:bgSim_002}
}
\centering
\subfloat[][]
{
\includegraphics[height=\gbsz]{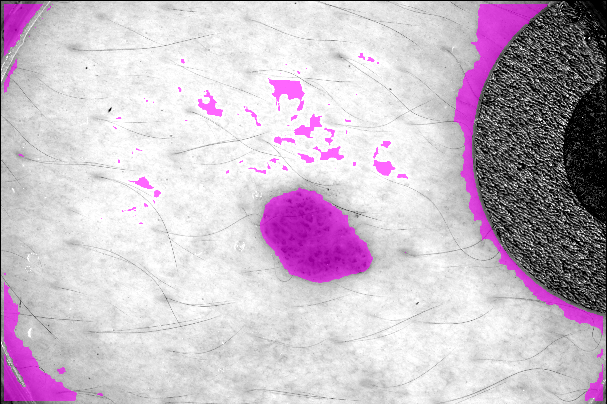}
\label{subfig:preSeg_002}
}
\centering
\subfloat[][]
{
\includegraphics[height=\gbsz]{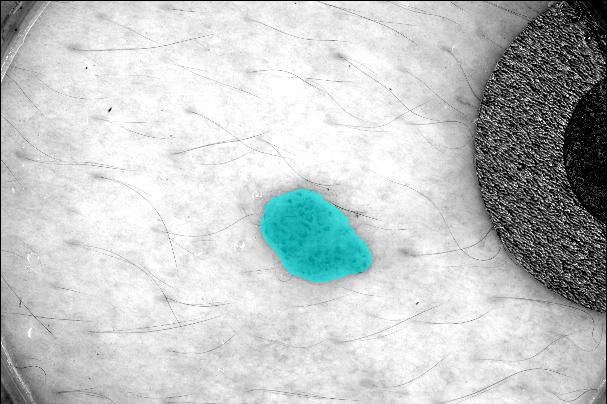}
\label{subfig:nevSegV0_002}
}
\centering
\subfloat[][]
{
\includegraphics[height=\gbsz]{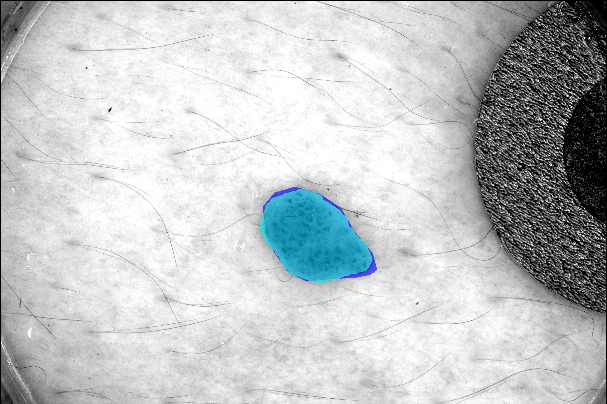}
\label{subfig:nevSegV0G_002}
}

\caption{Illustration of segmentation procedure in Specification~\ref{alg:seg} of image ISIC\_0008294 (also shown in Fig~\ref{fig:nevi} (e)). 
Figure (a) shows the nevus intensities (greyscale);
Figures (b) to (h) are each associated with a specific formula, as indicated below, with the exception of (d) 
where the score is shown as a varying intensity of yellow: (b) {\sf blackBorder} (red); (c) {\sf bgSample} (blue); (d) {\sf bgSimScore} (yellow); (e) {\sf bgSim} (green); (f) {\sf preSeg} (magenta); (g) {\sf nevSegV0} (cyan); (h) {\sf nevSegV0} (cyan) and ground truth (blue).
}
\label{fig:segsteps}
\end{figure}

\paragraph{Brief description of the \imgqool segmentation Specification~\ref{alg:seg}.} After importing the standard library ({\sf stdlib.imgql}), containing derived \imgqool operator definitions, and loading the image with ground truth and the related nevus image (lines 1-3), two quantitative images are defined: {\sf nevusImgIntens} for the nevus image and {\sf groundIntens} for the ground truth image (lines 4-5). These quantitative images associate to each voxel its intensity (luminosity); intuitively, each of them can be thought of as a (2D) matrix. 

In line 6 a similarity operator is defined with parameters {\sf a} and {\sf rad}, that is based on the cross correlation operator;   {\sf a} defines the sample area (denoted by $\Phi$ in Section~\ref{sec:SpatialLogicFramework}) and  {\sf rad} defines the radius around each voxel $x$ for the construction of the local histogram of $x$. We recall that the parameters of the cross correlation operator are, from left to right, the radius, the attribute of the voxels around $x$,
the attribute of voxels of the sample area, the minimum and maximum value of the intensity of the whole nevus image, and finally the number of bins in both histograms.

The (intermediate) results of the segmentation procedure defined in the rest of the specification are illustrated in Fig.~\ref{fig:segsteps}.
Lines 7-8 specify the characteristics of voxels that are part of the black corners that can be observed in many images (in a similar way as shown in Fig.~\ref{subfig:02} and Fig~\ref{subfig:04}). Such voxels should not be considered in the sample of the skin texture. In line 7 voxels are specified that are almost black, i.e. having an intensity below 40. Then (line 8) only those almost black voxels are considered from which the border can be reached exclusively `passing by' further almost black voxels, exploiting the {\sf grow }operator.

In lines 9-10 a sample ({\sf bgSample}) of the skin around the nevus is specified, namely a sample of voxels that are most likely part of the healthy skin without (or with very few) voxels that are part of the nevus. This sample consists of voxels that are at most at relative distance 200 ({\sf bgSampleWidth}) from the black border (lines 8-9). In line 11 the similarity score of each voxel in the image w.r.t. the sample is computed using the {\sf similarTo} operator and saved as a quantitative image. Line 12 characterises  all voxels that have a cross correlation score larger than 0.05. This line has been inserted only for illustration purposes here to highlight the voxels with skin texture; in Figure~\ref{subfig:bgSim_002} we show this set of points ({\sf bgSim}). 
%

A preliminary segmentation is specified in line 13, where we  look for voxels that are not part of the border and that are surrounded by voxels with a cross correlation score of less than 0.11, a small correlation score. The idea is that, at the border of the nevus and the healthy skin, the histograms of the area around those voxels represent in part the skin and in part the nevus, which have in general rather different intensity distributions. The cross correlation of such histograms with the sample area of the skin can therefore be expected to be quite small. 
The exact value of the threshold has been established in an empirical way; it is the value that gives on average good results for the investigated datasets. For optimal results on individual images this threshold value may differ slightly. 


Of course, this pre-segmentation should exclude areas close to the black border and in patches. The latter are used in some images to indicate the position of nevus with little contrast (see for example Fig.~\ref{subfig:1191} and Fig.~\ref{subfig:5787}). There may also be other small darker areas on the skin that are not part of the nevus. Therefore, in line 14, the preliminary segmentation is refined by taking only the largest volume ({\sf maxvol}) far enough from the black border and smoothening the specified area removing small noise and irregularities at the edge. 

In line 15 the same procedure of line 14 is repeated with a larger smoothening factor. This is used to exclude possible thin protrusions attached to the segmented nevus that are originating from thin hairs or shadows. In line 16 the intersection of these intermediate results is taken to preserve the more detailed edge of the nevus and at the same time to exclude some larger protrusions (i.e. several hairs grouped together). Also voxels that are part of possible patches are excluded from the segmentation and we obtain a first nevus segmentation in line 17 ({\sf nevSegV0}).

\begin{algorithm}
    \small
    \tt
\SetAlgoLined
\caption{\label{alg:res} Generating model checking results and similarity scores}
\imgImport
let groundTruth = groundIntens >. 0\\
save "\$OUTPUTDIR/\$NAME\_nevSegV0.png" nevSegV0\\
save "\$OUTPUTDIR/nevSegV0.nii.gz" nevSegV0 \\
print "DICE V0" dice(nevSegV0,groundTruth)
\end{algorithm}

In Specification~\ref{alg:res} the manual segmentation performed by domain experts (the `ground truth') is defined as a predicate that is satisfied by voxels in the image of the ground truth where the intensity of the voxel is positive (line 1). In fact, the ground truth is a black and white image of the same size as the image of the nevus where the area indicated by the expert is white (intensity 255) and the rest black. The resulting segmentation (but also other intermediate results as those shown in Fig~\ref{fig:segsteps}) can be saved in .png format or in the NIfTI (.nii) format. The latter format is used by various viewers used in medical imaging. We used the free viewer MRIcron\footnote{{\sf https://www.nitrc.org/projects/mricron}}.
The operator {\sf dice } compares (line 4) the segmentation defined by {\sf nevSegV0 } with the {\sf groundTruth }giving as result a similarity score as defined in Specification~\ref{alg:indexes}. Further details on these scores are provided in the next section.



\begin{algorithm}[t!]
    \small
    \tt
\SetAlgoLined
\caption{\label{alg:dist} Relative distances}
let refImgPerimeter = 2 .*. (1022 .+. 767)\\
let imgSizeFactor = (volume(border) ./. refImgPerimeter)\\
let relDist(x) = (imgSizeFactor .*. x)
\end{algorithm}

Specification~\ref{alg:seg} uses the {\sf relDist} operator defined in Specification~\ref{alg:dist}. The ISIC 2016 datasets contain images of very different sizes. The {\sf relDist} operator is introduced to scale the distance appropriately, with respect to a reference image.
The size of the reference image is defined as the length of its perimeter, i.e. the number of voxels on its border. The reference image has a width of 1022 voxels and a height of 767 voxels. The perimeter of the image being analysed can be found as the volume (i.e. number of voxels) that form the border (i.e. one voxel wide edge) of the image. The property {\sf border} is a built-in operator of \SLCSMI{.} The scaling of the distance is the fraction between the length of the perimeter of the image under analysis and that of the reference image.

\begin{algorithm}
    \small
    \tt
\SetAlgoLined
\caption{\label{alg:patch} Patches}
let bNev = blue(nevus)\\
let rNev = red(nevus)\\
let gNev = green(nevus)\\
%
let patchBlue = distleq(relDist(5),(bNev > (rNev +. 30)) \&\\ (bNev > (gNev)) \& (bNev >. 150)) \\
let patchRed = distleq(relDist(5),(rNev > (bNev +. 100)) \&\\ (rNev > (gNev +. 20))) \& (rNev >. 130) \\
let patchGreen = distleq(relDist(5),(gNev > (rNev +. 20)) \&\\ (gNev > bNev) \& (gNev >. 100)) \\
let patchPart(x,y) = ifB(volume(x) .<. (y .*. volume(tt)),x,ff) \\
let patchSample = patchPart(patchBlue,0.4) |  patchPart(patchRed,0.4) | patchPart(patchGreen,0.4) \\ 
%
let patchAtBorder = touch(smoothen(patchSample,relDist(10)), distleq(relDist(20),border)) \\
%
let patch = ifB(ppM(patchAtBorder) .>. 0.5, patchAtBorder, ff)
\end{algorithm}

Specification~\ref{alg:seg} also uses the predicate {\sf patch} that is satisfied by voxels that are part of a patch. Patches are defined in Specification~\ref{alg:patch}. Lines 1-3 define three quantitive images (matrixes) projecting the intensity of the blue, red and green part of the rgb-vector for each voxel of the image. Lines 4-9 define blue, red and green patches, respectively. These also cover intermediate hues such as yellow and orange. However, it is not enough to define the colour ranges of patches because nevi or skin may have occasionally colours in those ranges too (see for example Fig.~\ref{subfig:02} and Fig.~\ref{subfig:04}). Using further knowledge about the relative spatial position of patches (they are at the border of the image), their relative size (covering not more than 40 percent of an image) and their compactness (their Polsby-Popper measure,  of compactness of a shape\footnote{Also known as ``Isoperimetric quotient''.}, {\sf ppM}, is at least 0.5), the specification {\sf patch} is given in line 13. 
{\sf ifB} is the boolean if-then-else construct of \imgqool. The definition of {\sf ppM } is shown in Specification~\ref{alg:indexes} (lines 6-8) in the next section.

\section{Results}
\label{sec:results}


In the literature on medical imaging several indexes are used to compare the similarity between two segmentations of the same image, in particular the similarity between the manual and automatic segmentation. Commonly used similarity measures are the Dice index, the Jaccard index and the accuracy index. These coefficients give a result between 0 (no similarity) and 1 (perfect similarity). Further measures are the sensitivity (fraction of true positives) and specificity (fraction of true negatives).  
For example a Dice index of around 0.9 is considered as indicating very good similarity. The Dice index (D) and the Jaccard index (J) are related: $ J = D / (2-D)$. In Specification~\ref{alg:indexes} these common similarity indexes are defined in \SLCSMI\ so that they can be calculated during the analysis. Their definitions should be self-explanatory recalling that the operator {\sf volume(x) }  gives the number of voxels that satisfy property {\sf x}. It must be noted though, that no unique `gold standard' for comparison exists because also manual expert markings have a considerable level of variability. In~\cite{CNPGHHS2017} it is reported that the average Jaccard index of agreement between 3 pairs of clinicians that each segmented a subset of 100 images was 0.786.



A large set of dermoscopic images is available from the ISIC gallery\footnote{ See: {\sf https://www.isic-archive.com/\#!/topWithHeader/onlyHeaderTop/gallery}}.
Furthermore, two datasets of dermoscopic images were made available for the ISIC 2016 Challenge\footnote{These datasets can be found at {\sf https://challenge.isic-archive.com/data}}. One set of 900 images for {\em training} purposes and a {\em test} set of 379 images for the final evaluation of the competing methods.

In Table~\ref{tab:res} we show the mean scores for {\sf nevSegV0} of Specification~\ref{alg:seg} and compare them with the best mean scores\footnote{These scores were obtained on the ISIC 2016 Challenge {\em test dataset}.} of the teams that participated in the ISIC 2016 Challenge Part 1 on nevus segmentation~\cite{GutmanCCHMMH16}. The prevalent techniques used by the teams in the ISIC 2016 Challenge were based on deep learning~\cite{CNPGHHS2017}. 

For the comparison in Table~\ref{tab:res} we have used two restricted subsets of images from the ISIC gallery
 that satisfy the criteria that were mentioned in Section~\ref{sec:segmentation}. {\em Mean first 10} concerns the first 10 images of the ISIC gallery
 (images ISIC\_0000000 to ISIC\_0000010) with the exception of image ISIC\_0000004 in which the nevus is touching the border. These 10 images have been used to develop the specification and to manually calibrate some thresholds. They are also part of the datasets for the ISIC 2016 Challenge, with the exception of image ISIC\_0000005. Furthermore, some other typical example images from the ISIC 2016 training set have been used, such as those shown in Fig.~\ref{fig:nevi}, to develop the part of the specification dealing with patches and some other aspects.

Subsequently we have analysed a larger set of images. {\em Mean first 50} concerns the first 50 images of the ISIC 2016 gallery
(ISIC\_000000 to ISIC\_0000050) omitting\footnote{Images 4, 26 and 33 violate criterium (1), images 31 and 50 violate criterium (2). For image 11, due to a technical issue, we had the wrong ground truth (namely that of image 00), image 24 has too low contrast.} images  4, 11, 24, 26, 31, 33 and 50. Also these images are part of the datasets of the ISIC 2016 Challenge.
%
%
Table~\ref{tab:res} shows the various scores for {\sf nevSegV0} for these restricted subsets. In the first line of the table the best scores for the ISIC 2016 Challenge are reported for the complete ISIC 2016 test set (see~\cite{RonnebergerFB15,SLD17}). Although the latter set is different and much larger, the scores we obtained for {\sf nevSegV0} for the small sets are in line with these other scores. This means that it is possible to obtain segmentations of good quality with our method. 
%
%
%
%
%
%
%
\begin{table}[t!]
\begin{center}
$
\begin{array}{| l | r | r | r | r | r |} \hline
                              & \mbox{Accuracy} & \mbox{Dice} &  \mbox{Jaccard} & \mbox{Sensitivity} & \mbox{Specificity} \\ \hline \hline
\mbox{Best ISIC 2016}	 & 0,953		&  0,91              &  0,843	        & 0,91	& 0,965\\ \hline
\mbox{V0: Mean first 10}	 & 0,969           &  0,92              &  0,855           & 0,96	& 0,971\\ \hline
\mbox{V0: Mean first 50}	 & 0,945	        &   0,90	         &  0,828	        & 0,89	& 0,980\\ \hline
\end{array}
$
\end{center}
\caption{Similarity scores of the nevus segmentation method {\sf nevSegV0 } shown in specification~\ref{alg:seg} compared with the best score on the test set of the ISIC 2016 Challenge~\cite{CNPGHHS2017}.}
\label{tab:res}
\end{table}





\begin{figure}[b!]
\def\gbsz{1.5cm}
\centering
\subfloat[][]
{
\scalebox{1}[-1]{\includegraphics[height=\gbsz]{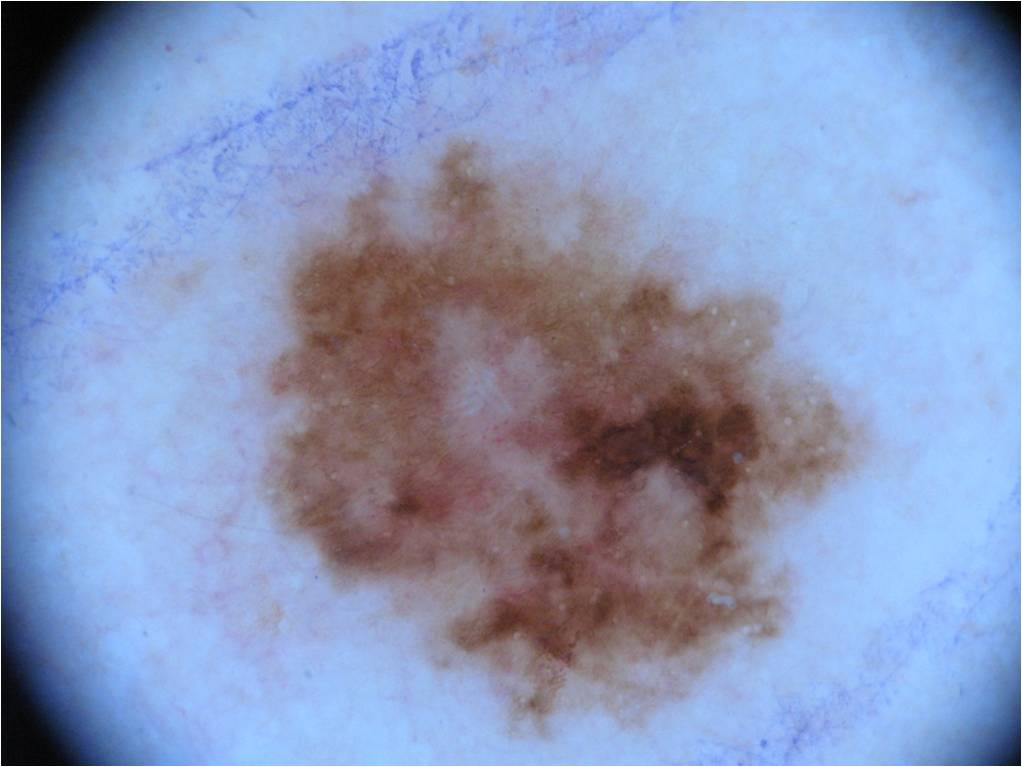}}
}
\centering
\subfloat[][]
{
\scalebox{1}[-1]{\includegraphics[height=\gbsz]{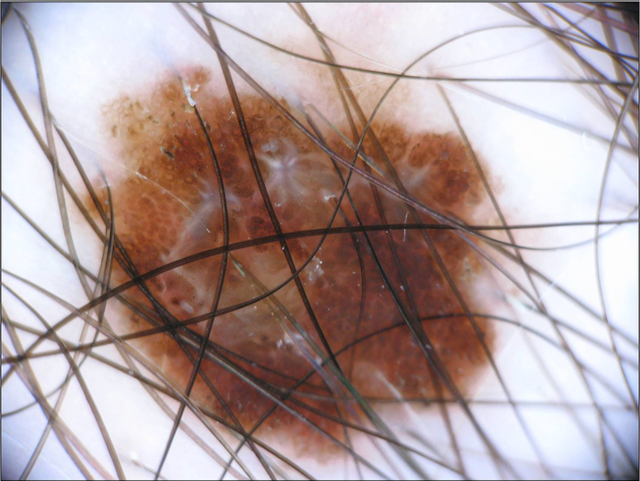}}
}
\centering
\subfloat[][]
{
\scalebox{1}[-1]{\includegraphics[height=\gbsz]{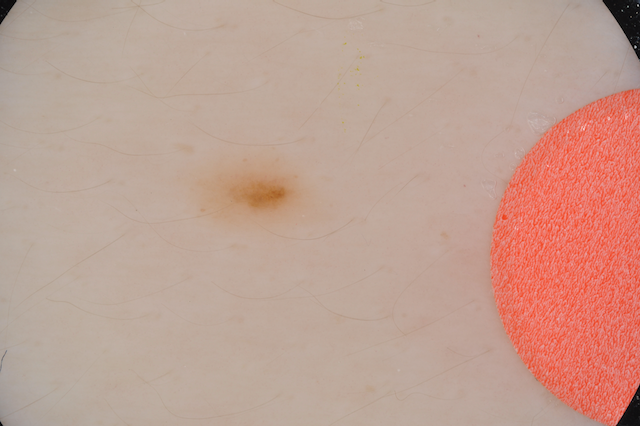}}
}\\
\centering
\subfloat[][]
{
\includegraphics[height=\gbsz]{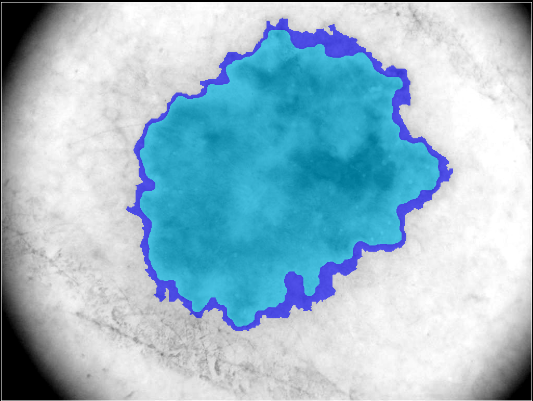}
}
\centering
\subfloat[][]
{
\includegraphics[height=\gbsz]{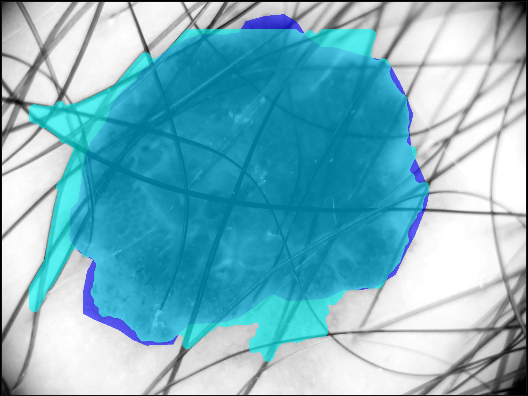}
}
\centering
\subfloat[][]
{
\includegraphics[height=\gbsz]{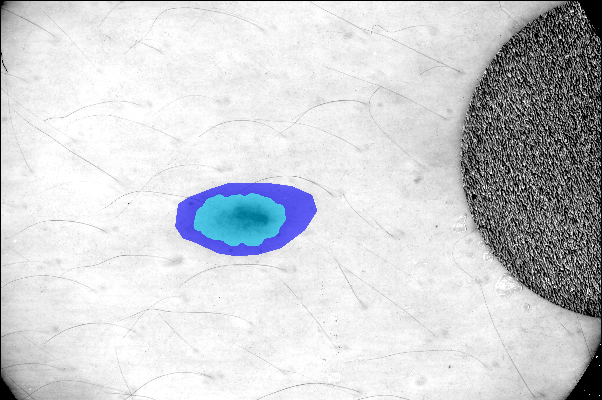}
}
%
\caption{Images and their segmentation (cyan) and ground truth (blue): ISIC\_0000002  (a) resp. (d), ISIC\_0000043 (b) resp. (e) and ISIC\_0004309 (c) resp. (f). }
\label{fig:segExam}
\end{figure}

\begin{algorithm}
    \small
    \tt
\SetAlgoLined
\caption{\label{alg:indexes} Similarity indexes and the Polsby-Popper measure}
%
let dice(x,y) = (2 .*. volume(x \& y)) ./. (volume(x) .+. volume(y))\\
%
%
let jaccard(x,y) = dice(x,y) ./. (2 .-. dice(x,y))\\
%
let sensitivity(x,y) = volume(x \& y) ./. (volume(x \& y) .+. volume((!x) \& (y)))\\
%
let specificity(x,y) = volume((!x) \& (!y)) ./. (volume((!x) \& (!y)) .+. volume((x) \& (!y)))\\
%
let accuracy(x,y) =  (volume(x \& y) .+. volume((!x) \& (!y))) ./. (volume(x \& y) .+. volume((!x) \& (!y)) .+. volume(x \& (!y)) .+. volume((!x) \& (y)))\\
%
 %
%
%
%
let square(x) = x .*. x\\
let iboundary(x) = near(interior(x)) \& !(interior(x))\\
let ppM(x) = (volume(x) .*. 4 .*. 3.14) ./. (square(volume(iboundary(x))))
%
\end{algorithm}
Next we investigate to what extent this is the case considering the much larger ISIC 2016 training set.
In Table~\ref{tab:dicedist} we consider the whole ISIC 2016 training set except those images where the nevus is touching the black border,
so in total we consider 750 of the 900 training set images. The table shows the number and fraction of images for which  {\sf nevSegV0} gives a Dice score above (or below) a certain threshold. This table shows that this rather simple specification gives an excellent score (Dice $> 0.9$) for more than 30\% of the images. Moreover, in more than 65\% of the cases it gives a reasonably good correspondence (Dice $> 0.7$). These results are rather surprising because of the enormous variability of the images and the relative simplicity of this version of the segmentation procedure. Fig.~\ref{fig:segExam} shows some more examples of segmentation via {\sf nevSegV0}. 
%

The illustration of the intermediate results of the segmentation procedure in Fig.~\ref{fig:segsteps} shows that with this spatial model-checking approach each step is `explainable', meaning that one can understand why a particular result is produced. This is one of the advantages of our spatial model-checking approach. It makes it much easier to further improve the segmentation procedure, to discuss a particular method with experts and to compare different methods or use them to document a particular analysis. Furthermore, the resulting segmentations are amenable to further analysis of the nevus area itself. For example one may extract further features such as regularity of its shape or texture, size and other features that may be of interest for the diagnosis of nevi. Note also that the segmentation method can be calibrated to specific images to obtain more precise results by tuning the thresholds.
We have not applied our method to the ISIC 2016 {\em test} set; we keep that set for  testing future versions of the specification.  
%
The results in this section have been obtained with \imgqool\  version {\sf 0.6.0\_osx-x64} on a MacBook Pro with 3.1 GHz Intel core i7 and 16GB of memory. For the large data sets an AMD Ryzen 7 2700 Eight-Core Processor with 32GB of memory has been used.
{\sf nevSegV0} takes approximately 2 seconds for a typical image of 1325 KB. The results together with the selected datasets discussed above are available in a git-repository\footnote{{\sf https://github.com/brocciagi/Spatial-Model-Checking-for-Nevus-Segmentation}}.

\begin{table}[t!]
\begin{center}
$
\begin{array}{| l | r | r |} \hline
\mbox{nevSegV0} &  \mbox{number} & \mbox{fraction (of 750)} \\ \hline
\mbox{Dice } >0.9 & 250  &  0.33 \\ \hline
\mbox{Dice } >0.8 & 396 &  0.53  \\ \hline
\mbox{Dice } >0.7 & 491  & 0.65  \\ \hline
\mbox{Dice } <0.5 &  140 &  0.19\\ \hline
\mbox{Dice } =0    & 45   &  0.06 \\ \hline
\end{array}
$
\end{center}
\caption{Dice score distribution of the nevus segmentation method {\sf nevSegV0 } for 750 images of the ISIC 2016 training set not touching the black border.}
\label{tab:dicedist}
\end{table}

\section{Conclusions and Future Work}
\label{sec:conclusions}


We have shown how spatial model checking techniques and the related tool \imgqool\ can be used for the segmentation of nevi. Nevus segmentation based on dermoscopic images is an important part of many automatic procedures to diagnose malign skin tumours such as Melanoma. To the best of our knowledge, this is the first time that spatial model checking is applied to this specific domain. Spatial model-checkers use high-level, often domain oriented, logical languages  to specify spatial properties. In this paper we have presented a segmentation method combining spatial operators inspired by the notion of closure spaces with more domain oriented operators such as a texture similarity operator. This first, rather simple method shows that an accuracy can be obtained that is in line with the state-of-the-art in nevus segmentation, i.e. a Dice score above 0.9. It has also been shown that it obtains this high accuracy in more than 30\% of 750 images of one of the largest available training sets of dermoscopic images that is publicly available. An advantage of this spatial model-checking method is that the segmentation procedure is explainable and high-level. This makes the method amenable to further improvements by inspection of the intermediate results, exchange and discussion of the method specifications between domain experts, conservation of the method for the purpose of documentation of the analysis and independent replication by other experts.

The  results we obtained so far are very promising and future work is envisioned to increase the class of images for which accurate segmentation can be obtained in a similar spirit as in which we have shown how one can deal with the presence of patches or the presence of other artifacts in the images that are due to the way the images have been produced. The enormous dishomogeneity in this type of images, both for what concerns the nevi and the images themselves, remains a great challenge.

\bibliographystyle{splncs03}
\bibliography{abbr.bib, bibliography.bib,xref.bib}

\end{document}